# Generating protein sequences from antibiotic resistance genes data using Generative Adversarial Networks

**Corresponding Author:** Prabal Chhibbar

Department of computer science and engineering

SRM Institute of science and technology

Tamil Nadu, Chennai-603203, India

1. Prabal Chhibbar

   Department of computer science and engineering,

   SRM Institute of Science and Technology

   Tamil Nadu, Chennai-603203, India

2. Arpit Joshi

   Department of computer science and engineering,

   SRM Institute of Science and Technology

   Tamil Nadu, Chennai-603203, India

# Introduction

In the last decade we have seen a boom in the number of antibiotics being developed and manufactured to cope and tackle with a variety of bacteria present or acquired throughout the human body. Bacteria are becoming resilient to these antibiotics rendering them ineffective [1]. This can be caused by the transfer leading to an exchange of genetic material through plasmids and transposons [2]. Bacteria like streptococcus pneumoniae [3] streptococcus pyogenes [4], and staphylococci [5] which cause an array of diseases like diarrhea, urinary tract infection and sepsis have become resistant to antibiotics. There are certain genes present in the genome of the organism which are responsible for this functionality of being resistant to an antibiotic [6] and it has now been proven that such genomes have an authoritative language model which can be used in natural language applications and can serve as viable input to natural language processing algorithms as well [7]. We introduce a method to generate synthetic protein sequences which are predicted to be resistant to certain antibiotics. We did this using 6,023 genes that were predicted to be resistant to antibiotics [8] in the intestinal region of the human gut and were fed as input to a Wasserstein generative adversarial network (W-GAN) model a variant to the original generative adversarial model which has been known to perform efficiently when it comes to mimicking the distribution of the real data in order to generate new data which is similar in style to the original data which was fed as the training data [9].The GAN architecture has been improved in order to cope with training instability in the form of the W-GAN model but recently as mentioned in Gulrajani et al, the architecture of the W-GAN model has been improved further to tackle failure to converge and low-quality samples, this was done by penalising the norm of gradient of the critic with respect to its input [10].We have applied the improved version of the W-GAN for our problem. The Wasserstein model of the generative adversarial networks has been employed before in context of generating genomic sequences. Wang et al. exploited the aforementioned architecture to further synthetic promoter design in the Escherichia coli model. 26 out of the 83 models were found to be functional and significantly expressed with varying activities, out of which 3 showed high promoter activity as compared to wild type promoters and their highly expressed mutants [11]. Killoran et al, designed and generated DNA sequences using deep generative models. They presented three approaches to do the same: generating DNA sequences using a generative adversarial network, a DNA based variation of the activation-maximisation method and a third approach based on the above mentioned approaches. They showed that such techniques capture important sequence information and applied the method to design probes for protein-binding microarray experiments to find that the sequences generated have far superior properties than the original sequences [12]. Besides, the generative adversarial network architectures various other architectures have been used in generating synthetic biomedical data an example being, of a deep reinforcement learning model which was trained and tuned to generate a query structure and compounds which are predicted to be active against a biological target, the model was even trained to generate analogues to the drug Celecoxib, this method could be extremely helpful in generating synthetic libraries from a single molecule [13]. On similar lines recurrent neural networks were used to generate novel molecules with significant affinity to biological targets, this method reproduced 14% of 6051 hold out test molecules in Staphylococcus aureus and 28% of 1240 test molecules that chemists designed in Plasmodium falciparum [14]. We hypothesize that by generating synthetic sequences using protein sequences of genes that are known to be associated with antibiotic resistance we can increase the scope of functionality of the genes that may be involved in the mechanisms involved in making a pathogen resistant to antibiotics and open new avenues for drug development and experimentation. We expect that the sequence that was generated by our model can be used to identify potential genes that can cause antibiotic resistance in a pathogen.

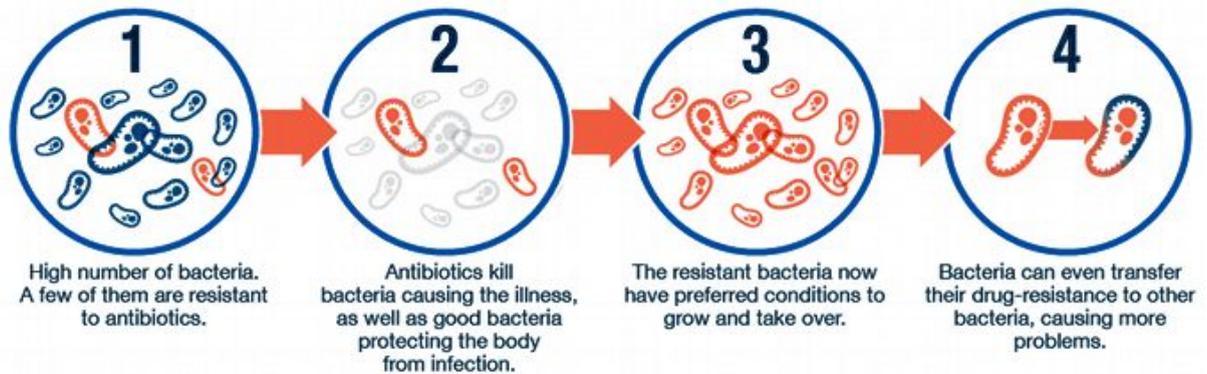

**Figure 1 - Steps occurred during antibiotic resistance**

## Methods

*Data Description*

We took 6,095 protein sequences of predicted and identified antibiotic resistance determinants (ARDs) taken from the human intestinal microbiota that were predicted using a new annotation modelling on a 3.9 million protein catalogue from the human intestinal microbiota as mentioned in Ruppe et al [15]. and present in the Mustard database as a downloadable FASTA file. (http://mgps.eu/Mustard/index.php?id=db)

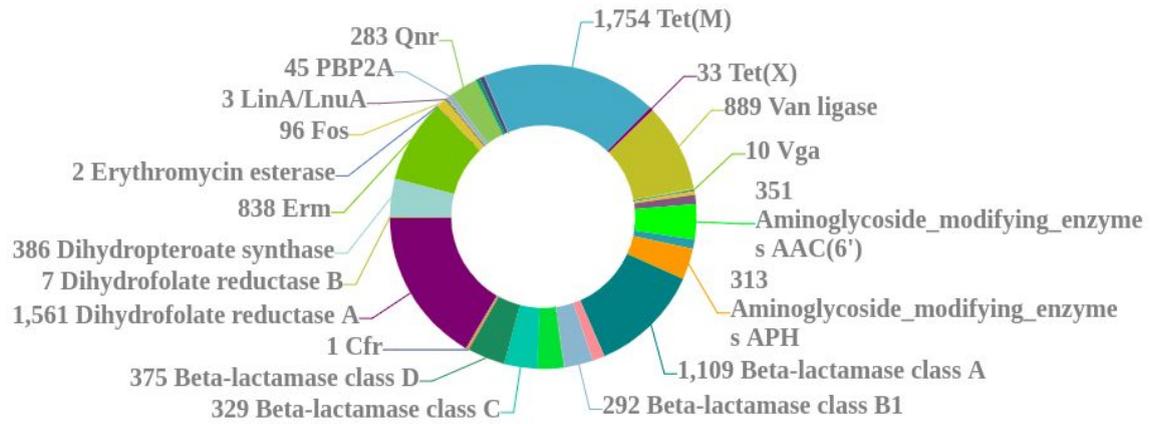

**Figure 2 - Statistics of the ARDs from the mustard DB, Source: MUSTARD DB**

**Encoding mode of promoter sequence**

Sequences with length N was represented as a numpy matrix. The length N in our scenario is set as 64. The encoding is performed as follows.

 a.  The sequence from the fastaa file is converted to a string
 b.  The string is then tokenized to unigrams
 c.  These unigrams are then encoded to unique integer values
 d.  Each of these integer values are converted to lists where the length of each list is the total unique characters present. All values are 0 except the value which the integer has. For ex - M = 8 (integer encoding) M = [0.0.0.0.0.0.0.1.0.0.0.0.0.0.]

This encoding process is better explained in the diagram below.

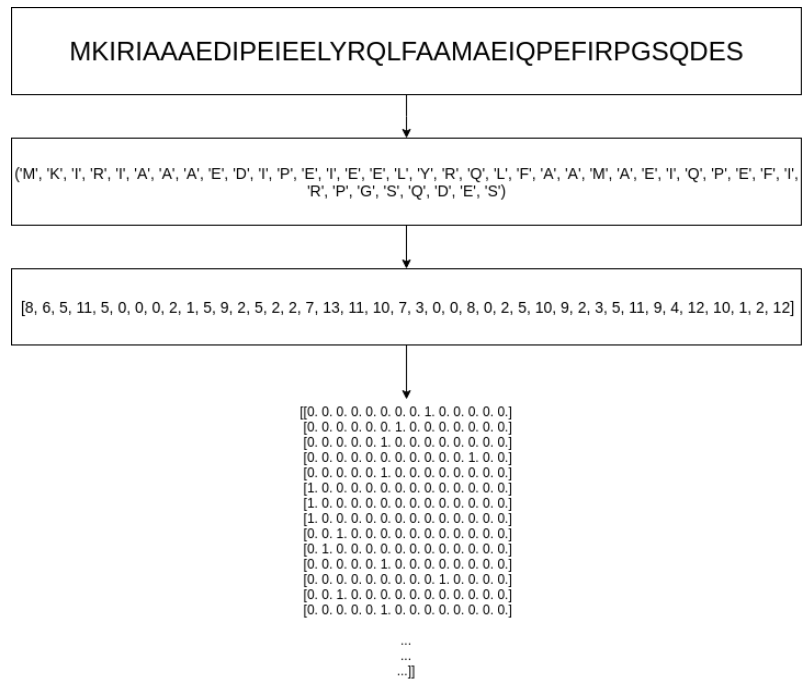

There might be other ways to encode the sequences such using a negative 1 for not present and 1 for present. The encoding method we use is generally used more often.

*W-GAN for generating sequences*

In our domain, the set of all possible data configuration is very large. There are many configurations of sub-sequences which are irrelevant or unlikely in real case scenarios i.e. non-motif characters. Hence, learning what sequences are relevant and what aren't is a very important feature for our model to learn. The disadvantage of using a single neural network on sequence data has been mentioned in Xueliang Leon Liu et al [16]. . Since the cost of computation is very high, a very large model cannot be created and since the data available is very less, the result has high accuracy but low recall.

A generative model in many scenarios has been successful in generating unique samples from a given training data. The goal of generative models is to capture the underlying realistic structure of the given data.

A generative network works essentially by training two networks together where these two networks are competing against each other. This was first done in Ian et. al [9]. There are two networks called a generator (generates samples from noise) and a discriminator (distinguishes between the training data and the generated samples). In our work, we use a specific type of generative adversarial network known as Wasserstein GAN mentioned in Martin et al [18]. This is based on the Wasserstein distance also known as the Earth mover's distance. GANs on their own sometimes do not properly converge on the most optimal solution. Some of the solutions to tackle this was mentioned in Arjovsky et al [19] where adding some noise to the generated data to stabilize the model.

Instead of adding noise, the WGAN model proposes a new cost function using wasserstein distance which has a smoother gradient everywhere.

As mentioned above, GANs contain two components (Fig. 1). First, a generator G transforms a continuous variable **z** into synthetic data **G(z).** It aims to generate new data similar to the expected one. The second component is a discriminator D, whose role is to distinguish generated data from real data.

Figure 3 - architecture of generator in GANs

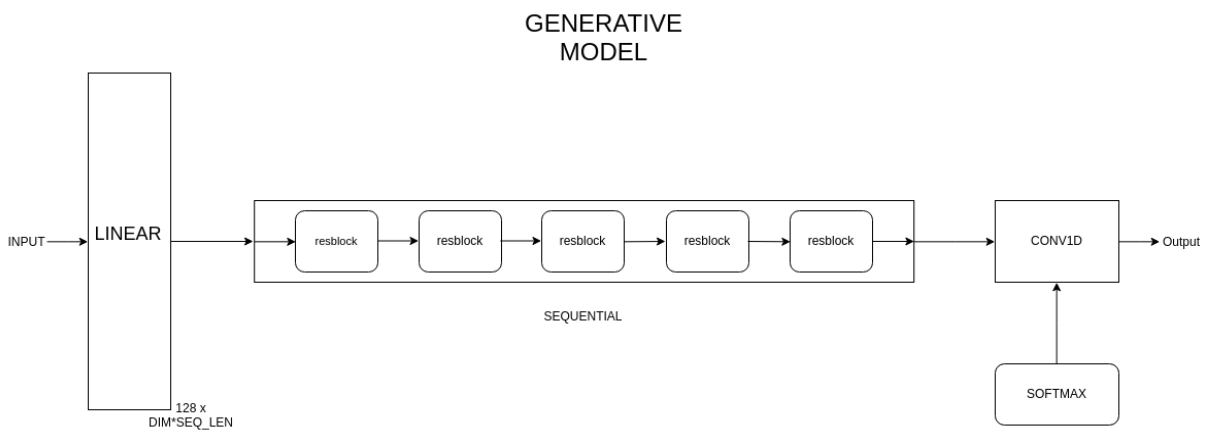

Figure 4 - Architecture of discriminator in GANs

ADVERSARIAL
MODEL

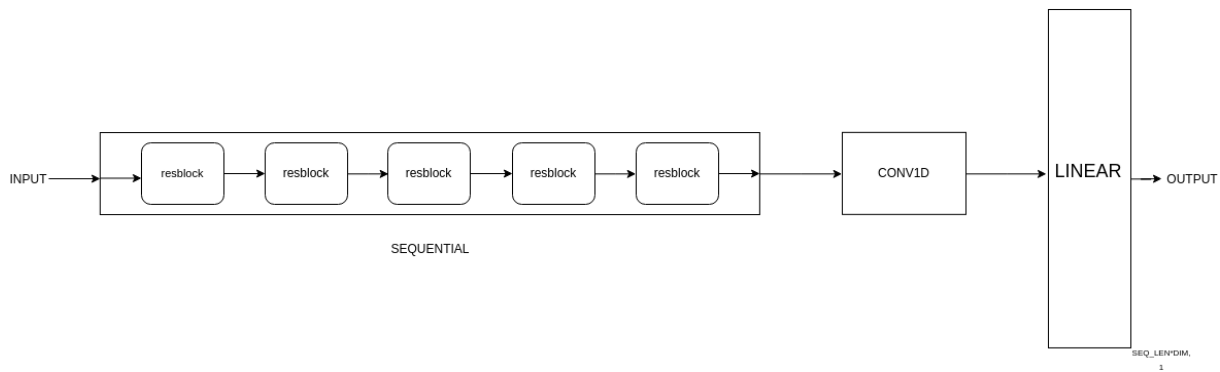

A neural network G(z,theta1) is used to model the generator. It's role is to map the random noise input z, into the desired data space (in our case, sequences). Conversely, the second neural network D(x,theta2) models the discriminator and outputs the probability that the data came from the real dataset in the range from [0,1].

As a result the discriminator is trained to correctly classify the input data as either real or fake. This means it's weights are updated as to maximize the probability that any real data input x is classified as belonging to the real dataset, while minimizing the probability that any fake image is classified as belonging to the real dataset. Therefore, the loss/error function maximizes the function D(x) and it also minimizes D(G(z)).

Since during training both the discriminator and generator are trying to optimize opposite loss functions, they can be thought two agents playing a minimax game with value function V(G,D). In this, the generator is trying to maximize its probability of having it's outputs recognized as real while the discriminator is trying to minimize this same value. Hence, the mathematical equation for the GANs network can be given as follows.

$$\min_G \max_D V(D, G) = \mathbb{E}_{\boldsymbol{x} \sim p_{\text{data}}(\boldsymbol{x})}[\log D(\boldsymbol{x})] + \mathbb{E}_{\boldsymbol{z} \sim p_{\boldsymbol{z}}(\boldsymbol{z})}[\log(1 - D(G(\boldsymbol{z})))].$$

We used the following training parameters to train our wgan.

| BATCH SIZE | 32 |
| ITERATIONS | 1000 |
| SEQUENCE LENGTH | 64 |
| DIMENSION | 512 |
| CRITIC ITERATIONS | 5 |
| LAMBDA | 10 |

**Table1: Parameters used for training the WGAN model**

*CARD Antibiotic resistance BLAST*

We provided our generated sequence as an input to the BLAST services provided by the comprehensive antibiotic resistance database(CARD: https://card.mcmaster.ca/analyze/blast) [21]. It performs a similarity search for mutations curated in the database. CARD is a bioinformatic database that consists of resistance genes, their phenotypes and associated products.

*BLASTp protein sequence alignment using NCBI BLAST*

We compared the test protein sequences and the generated protein sequences using the BLAST services provided by the National centre for biotechnology information (NCBI) [20]. This was done in order to check for the level of similarity between the two sequences.

*K-mer frequency counting*

We counted the frequency of the k-mers for values of k=2,3,4,5,6. The implementation of the same was written in Python 2.7. We found the common k-mers between the original sequence and the generated sequence. The frequency of the mentioned k-mers were plotted for comparison.

# Results

*Generated sequence using the W-GAN model: ARD-KATNISS*

We got the best results after 1000 iterations. We set the batch size of 32, sequence length of 64, dimension of 512 along with 5 critic-iterations which is the number of iterations the discriminator is trained for every one iteration of the generator. We used WEBLOGO [21] a web server based application to graphically represent biological sequences to represent the generated sequence. We dubbed the generated sequence as ARD-KATNISS. The generated sequence is available at:

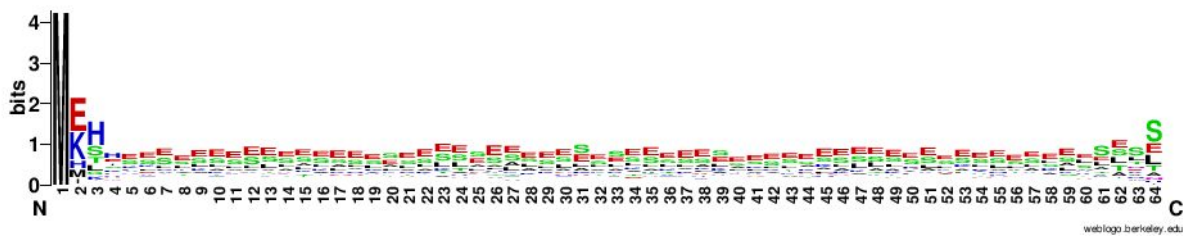

**Fig 5: The generated sequence after 1000 iterations represented using WEBLOGO**

*CARD BLAST detected VANs gene cluster in the generated sequence*

We ran the BLAST service provided by the comprehensive antibiotic resistance database (CARD). We got a hit for glycopeptide resistance gene cluster: VANs with the sequence matching closely to one found in Enterococcus gallinarum. The resistance of the aforementioned species to this particular class of antibiotics has been documented by Arias CA, et al [22].

**Results from BLASTp protein sequence alignment**

We compared the protein sequence as described in the data description section which was used as the training data and the protein sequence that was generated using the W-GAN architecture. We got an e-value of 6.5 and the measure of the similarity between the sequences in percentage was found out to be 27.88.

*K-mer frequency counting*

We counted the frequency of the k-mers for values of k=2,3,4,5,6. The implementation of the same was written in Python 2.7. We found the common k-mers between the original sequence and the generated sequence. The frequency of the mentioned k-mers were plotted for comparison. We found that the frequency of the k-mers were consistent for most of them and most of the k-mers were expressed.

| K-Value | Sequence | Total number of K-mers | Number of unique kmers |
|---|---|---|---|
| 2 | Sequence 1 | 398 | 244 |
| 2 | Sequence 2 | 244 | 244 |
| 3 | Sequence 1 | 7879 | 1866 |
| 3 | Sequence 2 | 1866 | 1866 |
| 4 | Sequence 1 | 143759 | 7964 |
| 4 | Sequence 2 | 7972 | 7964 |
| 5 | Sequence 1 | 1009900 | 14156 |
| 5 | Sequence 2 | 19303 | 14156 |
| 6 | Sequence 1 | 1649133 | 3934 |
| 6 | Sequence 2 | 29222 | 3934 |

**Table 2 : The number of unique K-mers found for different values of K, where sequence 1 is the original protein sequence and sequence 2 is the generated sequence**

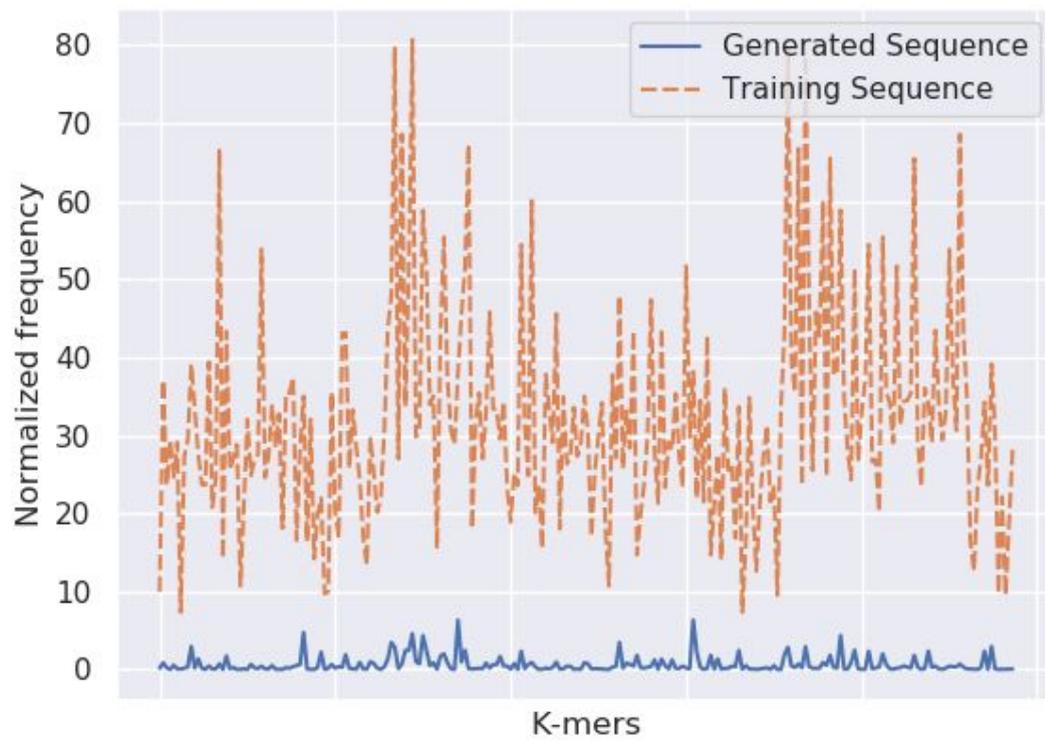

**Fig 6: K-mer analysis of generated vs training sequence where k=2**

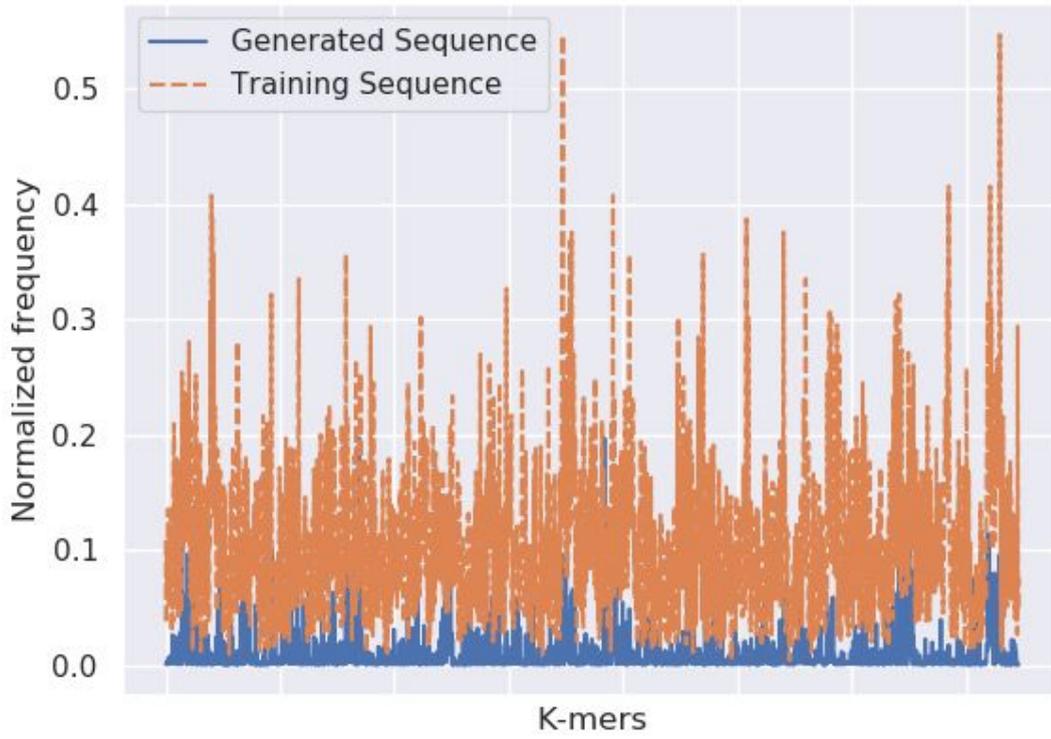

**Fig 7: K-mer analysis of generated vs training sequence where k=3**

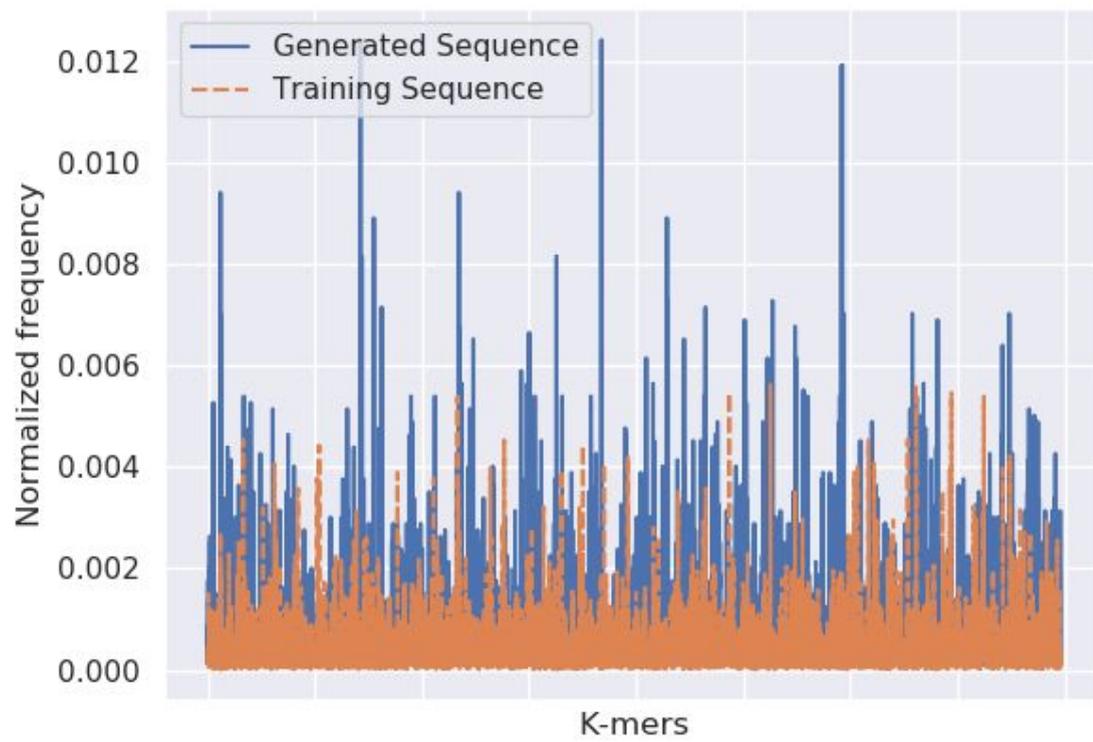

**Fig 8: K-mer analysis of generated vs training sequence where k=4**

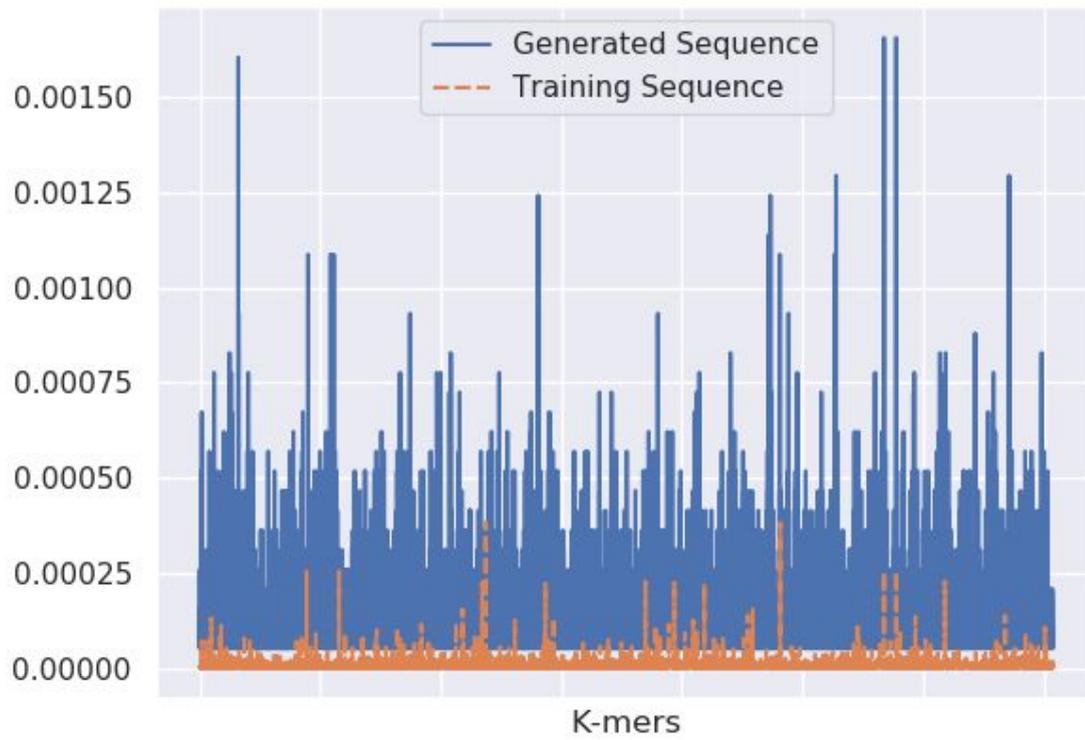

**Fig 9: K-mer analysis of generated vs training sequence where k=5**

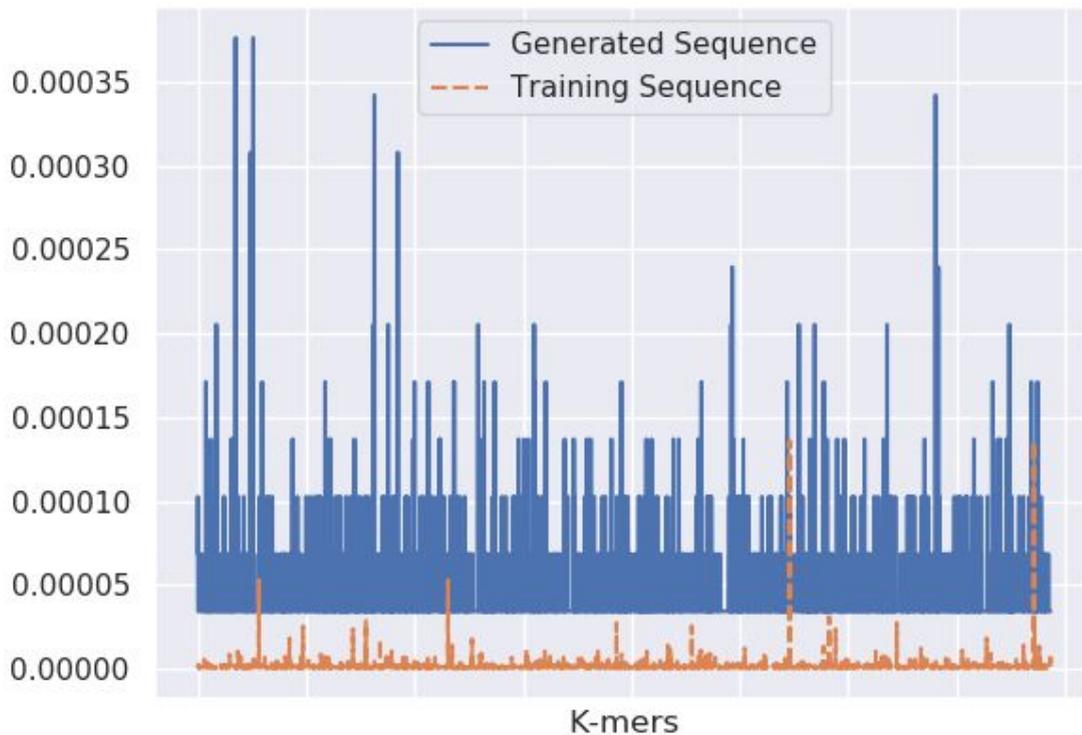

Fig 10: K-mer analysis of generated vs training sequence where k=6

## Discussions and conclusion

Intrinsic antibiotic resistance is a phenomenon, where the bacteria has an internal mechanism which makes it resistant to a particular class of antibiotics. This could be due to outer membrane permeability, enzymes or efflux pumps [23,24,25] amongst other reasons and mechanisms present in the concerned organism. We use the data which is associated with the commensal bacteria in the intestinal region of the human gut. We found that the WGAN model was able to generate the protein sequences and we got the best performance after a 1000 iterations. The generated sequence was then compared with the original sequence using the BLASTp alignment tool and was only approximately 29 percent similar to the original sequence showing that the model generated a novel sequence which can be used to study and expand functionality associated with the ARDs present in the gut region. We ran the CARD BLAST on the generated sequence where we got a hit with the VanS cluster showing resistance towards glycopeptides antibiotics in Enterobacteria gallenerium and opens avenues for the possibility of more such strains which have not been well documented and not present in the CARD database or any other source it uses for it's search. When we counted the k-mer frequency, we found that a lot of k-mers were common between the original and the generated sequence, showing that the model could take the structural properties of the sequence into consideration. It sometimes did under or overestimate the importance of certain k-mers but overall maintained consistency with the original sequence.

Generating sequences can open a whole area of study, where you simulate functionality where it is expensive to do the same in a wet lab experiment. With the case of antibiotic resistance, new information comes to light everyday and there are a lot of gaps in our knowledge which could be filled up by generating sequences